\documentclass[preprint]{aastex}
\usepackage{epstopdf}
\def\apgt{\ {\raise-.5ex\hbox{$\buildrel>\over\sim$}}\ }
\def\aplt{\ {\raise-.5ex\hbox{$\buildrel<\over\sim$}}\ }
\def\Msun{M_\odot}
\def\Lsun{L_\odot}
\def\Rsun{R_\odot}
\usepackage{epstopdf}

\shorttitle{Extreme Classical Novae}
\shortauthors{Shara et al.}

\begin{document}

\title{An Extended Grid of Nova Models. III. Very Luminous, Red Novae}

\author{
Michael~M.~Shara\altaffilmark{1}
Ofer~Yaron,\altaffilmark{2}
Dina~Prialnik,\altaffilmark{2}
Attay~Kovetz,\altaffilmark{2,3}
and
David~Zurek\altaffilmark{1}
}

\altaffiltext{1}{Department of Astrophysics, American Museum of Natural
History, Central Park West and 79th street, New York, NY 10024-5192}
\altaffiltext{2}{Department of Geophysics and Planetary Sciences,
Sackler Faculty of Exact Sciences, Tel Aviv University, Ramat Aviv
69978, Israel}
\altaffiltext{3}{School of Physics and Astronomy, Sackler Faculty of
Exact Sciences, Tel Aviv University, Ramat Aviv 69978, Israel}

\begin{abstract}

Extremely luminous, red eruptive variables like RV in M31 are being suggested as exemplars of a new class of astrophysical object. Our greatly extended series of nova simulations shows that classical nova models can produce very red, luminous eruptions. In a poorly studied corner of 3-D nova parameter space (very cold, low-mass white dwarfs, accreting at very low rates) we find {\it bona fide} classical novae that are very luminous and red because they eject very slowly moving, massive envelopes. A crucial prediction of these nova models - in contrast to the predictions of merging star (``mergeburst") models - is that a hot remnant, the underlying white dwarf, will emerge after the massive ejected envelope has expanded enough to become optically thin. This blue remnant must fade on a timescale of decades - much faster than a ``mergeburst", which must fade on timescales of millenia or longer. Furthermore, the cooling nova white dwarf and its expanding ejecta must become redder in the years after eruption, while a contracting mergeburst must become hotter and bluer. We predict that red novae will always brighten to $L \sim 1000 \Lsun$ for about 1 year before rising to maximum luminosity at $L \sim 10^{6} - 10^{7}\Lsun$. The maximum luminosity attainable by a nova is likely to be $L \sim 10^{7}\Lsun$, corresponding to $M \sim -12 $. In an accompanying paper we describe a fading, luminous blue candidate for the remnant of M31-RV; it is observed with HST to be compatible only with the nova model.

\end{abstract}


\section{INTRODUCTION}
\label{introduction}

The realization that classical novae are colossal explosions of previously nondescript stars was hammered home to the astronomical community with the appearance of first-magnitude GK Persei --- Nova 1901 A.D. Outshining all but a handful of stars in the night sky, it jump-started the scientific study of classical nova eruptions \citep{pay57}. The extraordinary light echoes \citep{rit01, cou39}, and ejected shells of matter \citep{rit17} associated with GK Per are the obvious hallmarks of a highly energetic, mass-ejecting event. The detection of erupting classical novae in the Small Magellanic Cloud \citep{mck51}, Large Magellanic Cloud \citep{gil27} and in M31 \citep{hub29} established novae as ubiquitous beacons radiating at tens of thousands of times the luminosity of the Sun or more --- a new class of astrophysical phenomenon \citep{har81}. Over the past century, of the order of a thousand erupting classical novae have been detected in the Galaxy, in galaxies of the Local Group \citep{cia90}, \citep{sha04} and as far away as the Virgo \citep{fer03, sha04} and Fornax clusters \citep{nei04}.

Less appreciated but just as important is that these early nova detections demonstrated the existence of close, interacting binary stars in galaxies outside our own. This could be appreciated only after two more seminal discoveries. First, \citet{wal54} demonstrated that nova Herculis 1934 (DQ Her) was a short period (4.65 hour) binary. A decade later, Kraft's \citep{kra64} spectroscopy demonstrated that old novae must be composed of a white dwarf (WD) accreting hydrogen-rich matter via an accretion disk from a red dwarf (RD) companion. Two logical consequences of this model are that (1) classical novae are powered by thermonuclear runaways (TNR) in their white dwarfs' hydrogen-rich envelopes \citep{sta72} and (2) novae must self-extinguish when their erupting envelopes are all but completely ejected \citep{pss77}.

Shortly afterwards it was realized that three parameters (WD mass, WD temperature or luminosity, and WD mass accretion rate) determine the physical characteristics of all nova eruptions \citep{sps80, Pri95}. Increasingly sophisticated input physics and rapidly increasing computer power led to self-consistent nova models including accretion, thermonuclear runaways (TNR) and ejection over many nova cycles. These culminated in the extensive grid of multi-cycle nova eruption models (\citet{pkz95}, hereafter Paper I). Covering what was then thought to be the entire 3-D parameter space that would produce novae, this set of models reproduced the range of energetics, timescales and mass ejections attributed at that time to classical novae \citep{war95}.

Some remarkable red, luminous eruptive variables are often claimed (e.g. in many of the papers in \citet{cor07}) to have physical properties inconsistent with all of the nova models in Paper I. While thermonuclear-powered nova eruptions can explain the luminosities of these objects, their massive and very cool ejecta seem totally different from those of other novae (\citet{mou90}, \citet{bon03}, \citet{mun02}, \citet{ban02}, \citet{kim02}, \citet{sok03}, \citet{kip04}, \citet{lyn04}). Does the nova model fail for these objects? Does this signal the existence of a hitherto unknown stellar eruptive phenomenon? Although V838 Mon appears to be incompatible with a nova model (as detailed below), are the other luminous red variables similarly constrained? This oft-repeated claim that M31-RV and similar, ultra-luminous red, eruptive variables cannot be explained as classical novae has prompted us to reexamine the 3-D parameter space that produces classical novae. The goal of this paper is to determine whether very luminous, red eruptive variables can, or cannot be produced by WD thermonuclear events. We also want to determine whether there are one or more observational signatures that can clearly differentiate between thermonuclear novae and other models of very luminous, eruptive red variables. If such red novae emerge naturally from our simulations then we might expect them to appear in nature. Of course, success in these efforts will not {\it prove} that thermonuclear novae are responsible for M31-RV or any given similar objects. But it will demonstrate that thermonuclear nova models should not be dismissed out of hand as the explanation for a very luminous red variable unless other evidence excludes such an interpretation. Only in the case of V838 Mon does such evidence seem to be in hand.

A pioneering attempt to consider very cold, low mass white dwarfs accreting hydrogen to produce very massive ejected shells was made by \cite{ibe92}. Such massive ejecta are essential to explain the cool red spectra. We have extended our grid of multi-cycle novae to lower white dwarf masses, lower accretion rates and colder white dwarfs than ever before, seeking the hard boundaries of nova 3-D space. We have found many new combinations that do give rise to classical novae, including those with the most extreme luminosities and ejected masses yet encountered. These models are described in \citet{yar05} (hereafter Paper II). Here we concentrate on two of those models and several additional ones that yield better fits to the basic characteristics of M31 RV than has been achieved before by any nova simulation. In an accompanying paper we report the discovery of a luminous, very blue star very close to the location of M31-RV that is behaving like the remnant of a classical nova eruption. Over the past decade that star has become much redder and fainter, and it now resembles old classical novae.

In $\S$~\ref{observations} we summarize the key observations that guide our new nova models. In $\S$~\ref{newmodels} we detail the initial models, and in  $\S$~\ref{Outbursts} we describe their evolution during nova eruptions. We compare the observed characteristics (detailed in $\S$~\ref{observations}) with the models in $\S$~\ref{confronting}, and offer testable predictions concerning M31-RV and the M85 eruptive variable. We briefly summarize our results and conclusions in $\S$~\ref{summary}.


\section{OBSERVATIONAL CONSTRAINTS}
\label{observations}

\subsection{M31 RV}

Determining accurate distances, and hence luminosities, for Milky Way stars is observationally challenging. It is no accident that classical novae and Cepheid variables were first calibrated as standard candles in the Magellanic Clouds and in M31. The well-established distance of the Andromeda galaxy makes the relatively well-studied system M31-RV (``RV" stands for red variable) the defining extragalactic member of its class. As pointed out by \citet{ric89}, at peak brightness M31-RV approached a bolometric magnitude $M_{\rm bol} = -9.6$, making it the most luminous red supergiant in the Local Group. The key observational constraints on this remarkable star have been presented by \citet{ric89}, \citet{mou90}, \citet{bry92}, \citet{tom92},  and \citet{bos04}. These observed characteristics and constraints (summarized below) must be mimicked by any model that seeks to explain this and similar outbursts.

\begin{enumerate}
\item The maximum observed brightnesses attained by M31 RV were V=15.3 \citep{bry92} and I=14.9 \citep{ric89}. The maximum absolute magnitude was thus at least $M = -9.3$, or $4\times10^{5}\Lsun$. The emission-line spectrum, together with the unknown dust absorption, precludes a definitive measure of the absolute magnitude at maximum. Using reasonable assumptions about reddening, M giant bolometric corrections and an adopted M31 distance modulus of 24.2, \citet{ric89} deduced $M_{\rm bol} = -9.96$, corresponding to $8\times10^{5}\Lsun$. The other authors noted above give values similar to this for the peak outburst luminosity of M31 RV.

\item The progenitor of M31 RV was fainter than I = 20.4 \citep{mou90}, thereby demonstrating an outburst amplitude exceeding 5.5 magnitudes.

\item The time to brighten the last 3 or 4 magnitudes to maximum light was less than 29 days, and could have been as short as hours or a few days \citep{bry92}.

\item M31 RV faded to $6\times10^{4}\Lsun$ at an epoch 70 days after maximum observed brightness. It faded below $10^{4}\Lsun$ 290 days after maximum brightness.

\item M31 RV displayed very red colors in the Gunn system (g-r = 1.33) and in the Johnson system (B-V = 1.9 and V-R = 1.0) near maximum light. At the same time it displayed a spectrum resembling that of an M0 supergiant. Infrared observations in the 1 to 3.6 micron range demonstrated that the ejected shell became as cool as 4000 Kelvins at maximum light, with a photospheric radius of $2000\Rsun$. About 70 days later the effective temperature dropped to 1000 Kelvins with a photospheric radius of $8000\Rsun$.

\item The ``coronal" or nebular phase of emission, seen in virtually all classical novae, did not appear in M31 RV. An ejected shell mass considerably larger than that of most novae therefore seems inescapable.

\item The velocity of ejection of material from M31 RV was $\sim150$~km~s$^{-1}$ \citep{mou90}. \citet{ric89} reported a +390~km~s$^{-1}$ radial velocity change in the variable over a span of eight hours, in both H$\alpha$ and stellar absorption lines. \citet{tom92} saw no such change, but did report significant [NII] emission lines flanking H$\alpha$, which might be rapidly variable.
\end{enumerate}

\clearpage
\subsection{The M85 Optical Transient}

This object is the most luminous of all the ``red novae".

\begin{enumerate}
\item At a distance of 15 Mpc the absolute R-band magnitude of this object at maximum was close to -12 \citep{kul07}. At 180 days after maximum it was producing $3\times10^{5}\Lsun$ \citep{rau07}.

\item This ``red nova" brightened by at least 7.7 magnitudes from its pre-eruption state.

\item The object faded most rapidly in blue light, and most slowly in the red. Nearly peak brightness was attained for 70 days, and the fading time thereafter was a few months.

\item Optical spectra show that the shell effective temperature was 4600 Kelvins near peak luminosity. Infrared multicolor images determine the ejecta's effective temperature to be 950 Kelvins at 6 months after the eruption began.

\item The optical material ejected from the M85 optical transient displayed a FWHM velocity of 350~km~s$^{-1}$, and an infrared expansion velocity of 870~km~s$^{-1}$.

\item Like M31-RV, the M85 optical transient (M85- OT) was not associated with any bright star forming region \citep{ofe08}. The g- and z-band absolute magnitudes of the progenitor were fainter than about -4 and -6 mag, respectively, corresponding to an upper limit for a progenitor (main sequence) mass of $7 \Msun$.

After this paper was completed a similar luminous red eruptive variable was discovered in the the Virgo galaxy M99 \citep{kas10}. Hubble Space Telescope archival imagery rules out red supergiants, blue supergiants and single main sequence stars earlier than type B2 as the progenitor.

\end{enumerate}

\subsection{V4332 Sgr}

The eruptive variable V4332 Sgr bears remarkable similarities to the luminous red objects we have already noted \citep{mar99}. There is not, unfortunately, a well-determined distance to V4332 Sgr, and hence its luminosity is quite uncertain. However, it is the only object of this class to have photometric observations in the years just before its eruption. A minimum fourfold brightening of V4332 Sgr in the decade before its eruption \citep{gor07} is an important constraint on models of this star.

\subsection{V838 Mon}

V838 Mon riveted the attention of astronomers when it erupted in 2002 January for two reasons. First, it nearly reached naked-eye brightness, peaking at V = 6.7 . Even more remarkable were the spectacular series of light echos surrounding the object. High resolution Hubble Space Telescope imagery and polarimetry of those echoes \citep{bon03} conclusively established the distance of V838 Mon at 6 kpc or greater. This in turn has allowed strong constraints on the physical characteristics of the variable, particularly its luminosity. At a distance of at least 6 kpc (and reddening E(B-V) = 0.8) the absolute magnitude of V838 Mon at maximum was at least $M_v = -9.6$ \citep{bon03}. V838 Mon and M31 RV are thus remarkably similar in their overall energetics, as well as their spectroscopic and photometric behaviors. The excellent distances available for each object have been crucial in demonstrating this remarkable agreement. The extremely red colors and cool spectra in the decline phase demand ejecta much more massive than has hitherto been predicted by nova models.

There is strong evidence that V838 Mon is associated with a B3V dwarf star, which is itself a member of a young association \citep{afs07} of at least 4 B-type stars. Such an association is too young to produce a cataclysmic binary, especially with a low mass white dwarf, unless strong dynamical stellar interactions and collisions occur \citep{ptz10}. (A possible formation mechanism would have the CO white dwarf core of a relatively massive star liberated from its envelope via collision with a neutron star, with the white dwarf capturing the neutron star's low mass companion in the process. But even then the time to cool the white dwarf, and to slowly accrete a massive envelope is much longer than the lifetime of a B3V star). Ordinary binary star evolution also cannot form a nova progenitor for V838 Mon within the main sequence lifetime of the B3V star.

There appear {\it not} to be correspondingly young stellar associations at the sites of M31-RV or M85-OT \citep{ofe08} or M99-OT, so that very luminous, red classical novae like those we describe below remain viable models for these events.

\section{NEW NOVA MODELS}
\label{newmodels}

In Paper II we published the results of evolutionary calculations of nova outbursts for a wide range of parameter combinations, spanning the 3-dimensional parameter space of novae: white dwarf mass $M_{WD}$, white dwarf core temperature $T_{WD}$ (correlated with the WD luminosity and hence age) and accretion rate $\dot{M}$. This grid of models was an extension to the grid published in Paper I ten years earlier. The parameter values adopted in the extended grid were: four $M_{WD}$ values---$0.65,1.00,1.25\ \&\ 1.40~M_\odot$, three $T_{WD}$ values---$10,30\ \&\ 50\ \times10^{6}$~K and eight $\dot{M}$ values---$10^{-6}$ through $10^{-12}~M_\odot$~yr$^{-1}$, as well as $5\times10^{-13}~M_\odot$yr$^{-1}$.

The hydrodynamic Lagrangian stellar evolution code used in all our studies is described in some detail in Paper I. It includes OPAL opacities \citep{igl96}, an extended nuclear reactions network comprised of 40 heavy element isotopes, and a mass-loss algorithm that applies a steady, optically thick supersonic wind solution (following the phase of rapid expansion). In addition, diffusion is computed for all elements, accretional heating is taken into account and convective fluxes are calculated according to the mixing length theory. Initial models were prepared for the four WD-mass values and three temperatures by cooling WD models from higher temperatures. Each nova model was followed through several consecutive outburst cycles. In each case, one cycle was then chosen as representative.
In Paper II we showed --- by analytical considerations supported by numerical calculation results--- that the parameter space where nova outbursts occur is limited. In order to test the extremes, we added calculations for still lower accretion rates ($5\times10^{-13}~M_\odot$~yr$^{-1}$), a few cases for $M_{WD}=0.4\Msun$, and several cases for $T_{WD}$ of only a few $\times10^{6}$~K. Many of these cases were found to lie outside the classical nova outburst parameter space; no eruptions ever occurred. Others presented features that appeared uncharacteristic of observed classical novae, but rather reminiscent of the properties of M31 RV. The results of two such extreme cases of the Paper II grid are given in Table~\ref{tbl:newgrid}, where $m_{acc}$ is the accreted mass, $m_{ej}$ -- the ejected mass, $Z_{ej}$ -- the mass fraction of heavy elements in the ejecta, $v_{av}$ -- the average expansion velocity, $T_{max}$ -- the maximal temperature attained in the burning shell at the base of the envelope, $L_{max}$ -- the peak luminosity, and $t_{3,bol}$ -- the time of decline of the {\it bolometric} luminosity by 3 mag. In the case of these very cold WDS, consecutive cycles are not identical, while they were identical for the hotter WDs considered. The differences between cycles are due to heating of the WDs by nova eruptions (see below). The results shown are for the first cycle, not as representative of all such eruptions, but as a reasonable example. We emphasize these models' unusual combination of very large ejected masses, low ejection velocities and very high (super-Eddington) peak luminosities.

M31 RV must have ejected a shell in excess of $10^{-3}\Msun$ to have become as cool as was observed. A {\it very cold} low-mass WD, accreting at a low rate is clearly the candidate to look for, if a very high ejected mass is sought, as was also pointed out by \citet{ibe92}. However, the 0.4$\Msun$ WD eruption declines too slowly to match the observations of M31-RV. We therefore decided to adopt an initial mass of 0.5$\Msun$, which is also more realistic for a C-O WD. We allowed the WD model to cool to core temperatures of $4\times10^{6}$~K ($3.4\times10^{9}$~yr), $3\times10^{6}$~K ($5.1\times10^{9}$~yr) and finally, $2\times10^{6}$~K ($8.8\times10^{9}$~yr). For each model we found the lowest accretion rate that still produced nova outbursts. For the middle $T_{WD}$ temperature, we also ran a model with a higher accretion rate, for comparison. Each model was evolved through several nova outburst cycles, as in Papers I and II. The results of these calculations are given in Table~\ref{tbl:charact}.

In contrast to the previous models, for which the WDs were hotter, the present, very cold models have very long cooling timescales, and thus the outer layers of the WD, which are heated during an outburst, do not cool back to the initial temperature before the next eruption. This is illustrated in Figure 1, where the new models are compared with a series of models from Paper II. Consequently, consecutive outbursts are not identical, but rather change monotonically, although slowly. Since the accreted mass decreases with increasing temperature of the outer layers, we list in Table~\ref{tbl:charact} the results for the first outburst in each run, which produces the highest ejected mass.

In Table~\ref{tbl:comp} we list abundances (by mass) of the main CNO isotopes, as well as isotopic ratios. We draw attention to the high oxygen content of the ejecta relative to carbon. This is typical of nova outbursts on low-mass WDs, as shown by \citet{kov97}; the trend is reversed for massive WDs.

Some of the parameters' trends reported in Tables 2 and 3 are not monotonic; the explanation is as follows. The nova phenomenon is a 3-parameter family of events  \citep{yar05}, and dependence on parameter values is not linear. For a given WD mass and accretion rate, $Z$ decreases with decreasing WD temperature. This is because more mass is accreted before the outburst when the WD is cold, while diffusion of hydrogen into the core at the inner boundary of the accreted shell slows down. As a result, the mixture of core material into the envelope (once convection sets in at ignition) is more diluted. This explains the differences between the first two rows in Table 2.

For given WD mass and temperature, $Z$ increases as $\dot M$ decreases because the accretion time is considerably longer and there is more time for diffusion. However, eventually, diffusion reaches equilibrium and effectively ceases and this is independent of $\dot M$. Therefore, if $\dot M$ decreases further, $Z$ will decrease due to dilution. This explains the difference between rows 4 and 2 in Table 2.  When two of the three parameters are changed at once, and these parameters each tend to change $Z$ in the opposite sense, it is even more difficult to predict the outcome. This explains why the results of line 3 in Table 2 may not be interpolated from the other results of the table.

In order to test the lowest possible core temperatures and their effect on the nova characteristics, we cooled three low-mass WD models for approximately a Hubble time ($1.3\times10^{10}$~yr) and then started accretion. At such low temperature an accretion rate of $10^{-11}\Msun$~yr$^{-1}$ did not produce a thermonuclear runaway (see Paper II for a discussion of the limited parameter space for TNR). We therefore adopted $\dot M= 10^{-10}\Msun$~yr$^{-1}$ for all these models. The results --- again, for the first cycle of the series --- are summarized in Table~4.

Generally, the results are similar --- high ejected masses, low velocities and very high peak luminosities --- although they differ in detail. Nova outburst characteristics are sensitive to even modest variations of one of the three basic parameters (see Papers I and II, and \citet{Sch94}).  Since small effects may arise from factors that are not included in the model (e.g. rotation or the presence of the secondary star) we do not aim for a precise match of any model with a particular ultra-luminous nova. That we are able to achieve good matches to the observations without any ``fine-tuning" speaks for the robust nature of the models.

The same arguments about non-motonically behaving parameters apply to Table 4. The mass accreted before outburst decreases with increasing WD mass. For a given accretion rate, one would expect higher $Z$ with increasing WD mass because the dilution is weaker, but also lower $Z$, because the time available for diffusion is shorter. Moreover, the larger the WD mass, the higher is the gravitational acceleration. This makes the diffusion of hydrogen inwards more difficult. It is not surprising, therefore, that the change of $Z$ with WD mass is not monotonic. Similar arguments apply to the ejection velocity. Larger WD masses imply higher degeneracy and stronger outbursts (with higher maximum temperature), which tend to lead to higher ejection velocities. The higher $g$, however, has the opposite effect on velocity, leading to the non-monotonic behavior of average ejection velocity with respect to the WD mass.

The strong outbursts and high luminosities of these very cold models are the direct result of the very high degeneracy attained at the base of the envelope just before the onset of the TNR. As an illustration, the Fermi parameter $\varepsilon_f\propto(\ln{P}-2.5\ln{T})$ ranges from 3.2 to 6.1 for the models in Table~2, whereas for the cold models ($T_{WD}=10^{7}$~K) of Paper II, the Fermi parameter is less than 2.

\section{MULTIPLE OUTBURST PEAKS: MULTIPLE ERUPTIONS OF ULTRA-LUMINOUS RED NOVAE}
\label{Outbursts}

The detailed evolution of the coldest $0.5\Msun$ WD model, accreting at a rate of $7\times10^{-11}\Msun$~yr$^{-1}$ (last but one entry of Table~\ref{tbl:charact}), is shown in Figures 2 through 7.

Intense nuclear burning (in highly degenerate matter) occurs for several {\it years} before an outburst, as shown in Figures 2, 3 and 4. A year before any detectable optical brightening, a strong thermonuclear pulse occurs at the base of the highly degenerate hydrogen-rich envelope, rising to $10^{10}\Lsun$,  and lasting only one day. This pulse releases enough energy to expand the white dwarf envelope a hundredfold. No subsequent pulse is as powerful, and we emphasize that this pulse, a full year before mass-loss commences, provides almost all of the energy needed to unbind the white dwarf envelope.

A very important characteristic of all these models is that {\it mass ejection is not continuous, but occurs in a few pulses}, separated by several contractions of the remaining envelope and subsequent rebound. This hydrodynamic-nuclear phenomenon was already pointed out by \citet{kov94}, and discussed in detail by \citet{PL95}. During the first mass-ejection producing pulse, illustrated in Figures 5 and 6, the mass of the nova envelope is sufficiently reduced that pressure at the envelope base drops significantly. This, in turn, causes the inner part of the envelope to start contracting, and pressure rises at the envelope base again. The enhanced pressure obtained in the burning shell at the base of the envelope produces a sufficient increase in nuclear luminosity to drive re-expansion of the inner envelope layers. The effect propagates to the surface, where a second pulse of mass loss occurs, sometimes characterized by higher expansion velocities. In some cases, a third, similar pulse occurs before the remnant envelope mass becomes sufficiently low, and the material sufficiently non-degenerate, to settle into quiet equilibrium burning until the remaining hydrogen is consumed. {\it Multiple peaks in the luminosity evolution of ultra-luminous red novae are a natural consequence of this expansion-contraction cycle}.

We discern three main episodes of mass loss, spanning about 150 days, preceded by a short and very mild episode some 50 days earlier. During each episode the velocity rises to a maximum value between 600-800~km~s$^{-1}$. The contraction of the WD at the end of each mass loss episode is shown by the sharp rise of the effective temperature on the one hand, and the rise in nuclear luminosity on the other. Expansion of the remnant shell follows, which is shown by the declining effective temperature (see Figure 7) and nuclear luminosity. At the end of the last mass loss episode we note a sharp drop of the nuclear luminosity, marking the end of the entire mass loss phase of the cycle.

Again in contrast to the Paper II models, characterized by higher core temperatures and/or WD masses, the outburst of the novae of Table~\ref{tbl:cold} occur in two stages. First, a sharp rise of the bolometric luminosity to about $10^{3}\Lsun$ lasts about one year. {\it We predict that this year-long ``pre-maximum" luminosity phase must occur in all ultra-luminous, red, thermonuclear-powered novae}. Only after a year does the luminosity rise to its maximum, accompanied by mass loss. The final rise in luminosity - spanning a few weeks - is of only about 6 magnitudes. The decline of the bolometric luminosity is also very slow, typical of low mass, cold WDs.  After the initial rapid expansion to about one solar radius noted above, the expansion proceeds at a slower rate. Significant mass loss commences only when the radius has increased to about $100\Rsun$, typical of nova outbursts. The initial expansion and rise in luminosity occur when the convective zone (starting at the base of the burning shell) extends all the way to the surface. The relatively slow expansion is due to the unusually large mass (and inertia) of the envelope as compared to the less extreme models of Papers I and II.

The very high mass-loss rate (of order $10^{-3}\Msun$~yr$^{-1}$) associated with ejecting the massive envelopes discussed here from a $0.5\Msun$ WD in just a few months would normally demand luminosities well in excess of $10^{6}\Lsun$, if those envelopes were promptly ejected from the WD surfaces. We emphasize that this is not the case. Mass loss only begins when the envelope has expanded to roughly $100\Rsun$. This bloated configuration arises because of the slow expansion phase, powered for about 1 year by a thermonuclear luminosity of $2\times10^{5}\Lsun$. This slow expansion phase largely unbinds the envelope, making subsequent mass-loss much easier to achieve with more modest luminosities.

\section{CONFRONTING MODELS WITH OBSERVATIONS OF M85-OT and M31-RV}
\label{confronting}

Soker \& Tylenda (2006) have compared and contrasted classical novae, born-again AGB stars and stellar mergers as possible models for very luminous, red eruptive variables. They considered multiple characteristics in their comparisons between models and theory: increase in luminosity factor, multi-outburst light curve, fading as a very cool supergiant, peak luminosity, solar ejecta abundances, outflow velocity, association with a young B3V star, and the circumstellar non-ionized matter illuminated by the outburst. They concluded that the born-again AGB model fails almost all of these comparisons, and we concur. They also concluded that classical novae fail on six of the nine criteria; we disagree, and detail below how, in fact, classical novae are compatible with all of the observations. Furthermore, we note that the merger model fails to explain the observed blue remnant for M31-RV \citep{sha11}, whose properties are summarized below.The nova model naturally predicts this observation.

\subsection{Mass of nova ejecta}

The ejected masses obtained in this study are the highest that have ever been predicted by the classical nova TNR scenario. We have shown that $2-3\times10^{-3}\Msun$ is the upper limit of $m_{ej}$ for classical novae. Suggestions of even larger masses - $10^{-2}\Msun$ or even $10^{-1}\Msun$ - have been made for M31-RV, based on the assumption of energy equipartition in the ejecta between kinetic energy and photons. All of the models of \citet{yar05} show that the equipartition assumption is incorrect \citep{sha10}. Radiated energy dominates kinetic energy by one to four orders of magnitude in virtually all nova eruptions. Nova ejecta masses based on the assumption of equipartition of energy are simply wrong, overestimating the ejected mass by at least an order of magnitude \citep{sha10}.

\subsection{Nova remnant}

Another critical prediction of the nova model is that ``red novae" must leave hot, blue remnant stars - cooling white dwarfs. We have used HST to search for such a hot, blue remnant at the site of M31-RV. In an accompanying paper we note the detection of just such an object, fading on a timescale of years, very close to the position of M31-RV \citep{sha11}. That blue object was observed to be radiating at least $10^{3}\Lsun$ when it was observed by HST seven years after M31-RV erupted. Twelve years after the eruption it was still radiating at $250 \Lsun$, and displaying an effective temperature of about 40 kKelvins. Twenty years after the eruption the luminosity of the object was still $10^{2}\Lsun$, and its effective temperature no cooler than 8kKelvins. The observed cooling and fading behavior of this blue remnant is in perfect accord with the nova models of this paper. The mergeburst model predicts a much cooler (3000 Kelvins) object that fades much more slowly (on a timescale of many nillenia) than is observed.

We predict that such a similar blue object must also eventually emerge from the the optically thinning ejecta in M85-OT if the nova model for that object is correct.

\subsection{Stellar Populations}

M31-RV erupted in the bulge of the Andromeda galaxy. \citet{bon06} examined the site of the eruption in Hubble Space Telescope images and reported that ``there is no evidence for any significant young population at this location in the M31 bulge".

M85-OT erupted near the lenticular galaxy (Hubble type S0) M85. \citet{kul07} examined the site of the eruption in HST, Spitzer Space Telescope and Chandra X-ray Observatory images and reported that ``There  is no evidence for a bright progenitor, and nor do we see tracers of massive star progenitors." These same authors summarized their search with ``We conclude that the M85-OT2006-1 probably arises from a population of stars with masses of a few times the mass of the Sun or less". The stellar populations around M31-RV and around M85-OT are thus consistent with those expected for classical novae.

\subsection{Oxygen}

A prediction of the models presented in this paper is that the ejecta in luminous red novae will be extremely oxygen rich, at the expense of carbon. The relatively cool peak temperatures and massive envelope of the low mass white dwarf we studied here allowed the CNO bi-cycle to ``cook" about 90\% of its initial C, N and O into oxygen and nitrogen.
\subsection{Carbon isotopes}

The nova models of this paper produced more Carbon-12 (relative to Carbon-13) than any we have ever computed. During the latter stages of ejection, C12/C13 ratios of 6 were encountered in the models. This should be compared with typical interstellar values of 60 to 90 \citep{mer79}.

\subsection{Luminosity increase}

An equally strong prediction of our low mass WD nova models is a pre-maximum rise to, and year-long plateau in luminosity at about $1000\Lsun$.  Neither M31-RV nor the M85 variable was observed long enough, or with sufficient sensitivity before maximum to know if it displayed this behavior. Multiyear synoptic surveys (such as the Palomar Transient Factory \citep{law09} and the Large Synoptic Survey Telescope \citep{lss09}) will surely find more such luminous, red eruptive objects. We predict pre-maximum outbursts, lasting about 1 year, with luminosities of about $1000\Lsun$ for the future luminous red eruptive variables that are extreme classical novae.

\subsection{Peak luminosity}

A challenging observation to explain for the classical nova scenario is the high claimed luminosity for the M85 transient. At peak brightness it was about 3 times larger than the most luminous super-Eddington nova models we have presented in this paper. However, it is important to note that the nuclear luminosities of model novae during flashes - including the models calculated here - exceed the apparent luminosities of even M85-OT by several orders of magnitude. Modeling the radiated luminosity depends critically on the expanding envelope opacity.

The OPAL opacities \citep{igl96} used in our simulations are only valid for temperatures greater than 6000 Kelvins. They don't include the effects of molecules that become important at lower temperatures. Temperatures below 6000 Kelvins have not been important in nova simulations until now because the much lower ejected masses of all previous models became optically thin with temperatures in excess of 6000 Kelvins. Recent low temperature opacity tables \citep{fer05} include the effects of hundreds of millions of atomic and molecular lines, grains, negative ions, and bound-free and free-free opacity. An important and remarkable result is summarized in Figure 11 of \citet{fer05}. The Rosseland mean opacity $\kappa$ decreases monotonically, by five orders of magnitude, as the temperature decreases from 10,000 to 3000 Kelvins. $\kappa$ then rises (almost) monotonically by four orders of magnitude as the temperature decreases from 3000 Kelvins to 700 Kelvins. Furthermore, $\kappa$ is a factor of 20 to 25 smaller at 3000 Kelvins than at 6000 Kelvins. This deep, well-defined opacity minimum has important consequences for nova envelopes - which we have encountered for the first time in the simulations presented here - that are massive enough to be optically thick at 3000-4000 Kelvins.

Massive nova envelopes, like the ones in the present simulations,  must ``leak" large amounts of radiation much faster than less massive nova envelopes whose effective temperatures always remain in excess of 6000 Kelvins. The maximum recorded luminosities of M31-RV and M85-OT were $8\times10^{5}\Lsun$ \citep{ric89} and almost $10^{7}\Lsun$ \citep{kul07}. These super-Eddington luminosities were observed when those objects displayed the remarkably red colors of early M supergiants, corresponding to effective temperatures of about 4000 Kelvins. While it is beyond the scope of the present paper to include a realistic model atmosphere with low temperature opacities in the simulations of low mass WD novae, we can quantitatively estimate the highest luminosity that any nova can attain. For a given WD mass Mwd, the luminosity L of a nova envelope radiating near the Eddington limit is given by equation 2b of \citet{sha81}:

$$ log(L/\Lsun) = 4.59 + log(M_{wd}/\Msun) +log(\kappa_{es}/\kappa) $$  where $\kappa$ is the envelope opacity and $\kappa_{es}$ is the opacity due to electron scattering. The key result is that the nova luminosity is inversely proportional to the opacity of its expanding envelope. At 3000 to 4000 Kelvins a nova envelope's opacity (as noted above) will be 20 to 25 times smaller than the opacity used in the simulations described above. We therefore estimate that the highest luminosity attainable by a nova like the ones we have simulated (which reach $3\times10^{5}\Lsun$) is roughly 25 times larger: i.e. $7.5\times10^{6}\Lsun$. This is very close to the luminosity observed for M85-OT. As we've already noted, there is more than sufficient energy generated (and ``bottled up" ) in the model envelopes to power a luminosity spike to $10^{7}\Lsun$. We suggest that it is the opacity minimum noted above that permits the rapid leak of energy corresponding to the energy spike observed in M85-OT.

\subsection{Fading as a very cool Supergiant and outflow velocities}

It is often noted that the very late spectral types (L or T) and cool temperatures of M31-RV and M85-OT are completely different from the behavior of post-outburst classical novae. This is because most classical novae eject shells of order $10^{-5}\Msun$ - at least a hundred times less massive than the ejecta of M31-RV-like objects must be to become as cool as is observed. Massive ejected shells ($10^{-3}\Msun$ ) must remain optically thick to radii 5-10 times larger than those of most classical novae. As noted above, their effective temperatures will cool to values considerably lower than those of classical nova blackbodies before they become optically thin - i.e. as low as 3000-4000 Kelvins, corresponding to the opacity minimum noted above. As these model shells are ejected at velocities of a few hundred km per second (10 times lower than those of most classical novae, and in accord with observations), they must remain optically thick for years instead of months, just as is observed.

\subsection{V838 Mon association with a young B3V star}

V838 Mon lies very close to (but is not coincident with) a B3V star in a young open cluster \citep{afs07}. \citet{tyl06} emphasize that there is not enough time during the lifetime of a B3V star to form the white dwarf needed for a classical nova eruption, and we agree (barring a very rare dynamical collision event). If the erupting object V838 Mon is a companion of the B3V star, or if it is a co-eval member of the young open cluster then it almost certainly cannot be a classical nova.

Remarkably, no such young stellar populations exist near M31-RV (and probably not M85-OT or M99-OT). A B star at the location of M31-RV would have been detected in archival HST images... it isn't there. Thus M31-RV cannot be ruled out as a classical nova.

\section{SUMMARY} \label{summary}

It is common to suggest that a few very luminous, red eruptive variables cannot be classical novae, and must therefore represent a new class of astrophysical objects.  In this paper (and in an accompanying one) we show that models of classical novae can, in fact, reproduce the observed phenomena rather well for M31-RV, M85-OT and the newly discovered red nova in M99 \citep{kas10} - but not v838 Mon. The underlying white dwarfs in such nova systems must be low in mass, cold, and accrete at a low rate from their companions. Predictions of the nova theory are in remarkably good accord with observations of the ``ultra-luminous red novae" M31-RV and M85-OT in terms of time scales and light curves.

The key question - are ``luminous red novae" really extreme classical novae or a new astrophysical phenomenon like mergebursts - will be settled when we detect more such objects well before maximum light, and when we recover more post-outburst remnants after the ejecta expand and become optically thin. The merger of two stars must swell the resulting star, leaving behind a cool, red remnant. The swollen remnant's envelope must contract, becoming hotter and bluer on a thermal timescale (of many millenia). This predicted behavior is in sharp contrast to that expected from a classical nova.

Classical novae also fade, but on a much faster timescale - years to decades. The underlying white dwarfs and accretion disks of old novae cool in the decades following an eruption, and thus become redder.  The object we have detected at the site of M31-RV attained a luminosity that is consistent with it being a classical nova. It is observed with HST to have been very hot about a decade after its eruption. It is fading on a timescale of decades, and becoming much redder as it does so. All of these observations are in accord with the nova models of this paper, but not with mergeburst models.


\begin{deluxetable}{ccccccccccc}
\tablecaption{Characteristics of The Outburst -- Paper II models\label{tbl:newgrid}}
\tabletypesize{\scriptsize}
\tablewidth{0pt}
\tablehead{
\multicolumn{3}{c}{Parameter Combinations} &
\multicolumn{7}{c}{Outburst Characteristics} \\
\colhead{$M_{WD}$} & \colhead{$T_{WD}$} & \colhead{log$\dot{M}$} &
\colhead{$m_{acc}$} & \colhead{$m_{ej}$} & \colhead{$Z_{ej}$} &
\colhead{$v_{av}$} &
\colhead{$T_{8,max}$} & \colhead{$L_{4,max}$} & \colhead{$t_{3,bol}$} \\
\colhead{($\Msun$)} & \colhead{($10^{6}$~K)} & \colhead{($\Msun$~yr$^{-1}$)} &
\colhead{($\Msun$)} & \colhead{($\Msun$)} & \colhead{} &
\colhead{$(km~s^{-1})$} &
\colhead{($10^{8}$~K)} & \colhead{$(10^{4}~L_\odot)$} & \colhead{(days)}\\
}
\startdata
0.40 & 10 & -11 & 5.9E-4 & 7.0E-4 & 0.21 & 460 & 0.95 & 23.9 & 5.5E4 \\
0.65 & 10 & -12 & 3.9E-4 & 6.7E-4 & 0.45 & 220 & 1.59 & 20.7 & 8.7E3 \\
\enddata
\end{deluxetable}

\begin{deluxetable}{ccccccccccc}
\tablecaption{Characteristics of The Outburst -- Models with $M_{WD}=0.5\Msun$\label{tbl:charact}}
\tabletypesize{\scriptsize}
\tablewidth{0pt}
\tablehead{
\multicolumn{3}{c}{Parameter Combinations} &
\multicolumn{7}{c}{Outburst Characteristics} \\
\colhead{$M_{WD}$} & \colhead{$T_{WD}$} & \colhead{$\dot{M}$} &
\colhead{$m_{acc}$} & \colhead{$m_{ej}$} & \colhead{$Z_{ej}$} &
\colhead{$v_{av}$} &
\colhead{$T_{8,max}$} & \colhead{$L_{4,max}$} & \colhead{$t_{3,bol}$} \\
\colhead{($\Msun$)} & \colhead{($10^{6}$~K)} & \colhead{($\Msun$~yr$^{-1}$)} &
\colhead{($\Msun$)} & \colhead{($\Msun$)} & \colhead{} &
\colhead{$(km~s^{-1})$} &
\colhead{($10^{8}$~K)} & \colhead{$(10^{4}~L_\odot)$} & \colhead{(days)}\\
}
\startdata
0.50 & 4 & 5E-11 & 1.5E-3 & 1.7E-3 & 0.12 & 271 & 1.38 & 21.8 & 2.6E3 \\
0.50 & 3 & 5E-11 & 2.0E-3 & 2.2E-3 & 0.10 & 329 & 1.44 & 20.8 & 4.2E3 \\
0.50 & 2 & 7E-11 & 1.2E-3 & 1.4E-3 & 0.13 & 336 & 1.34 & 26.1 & 2.7E3 \\
\hline
0.50 & 3 & 1E-10 & 6.7E-4 & 7.4E-4 & 0.14 & 104 & 1.38 & 17.9 & 2.0E3 \\
\enddata
\end{deluxetable}

\begin{deluxetable}{ccccccccc}
\tablecaption{Ejecta Composition -- Models with $M_{WD}=0.5\Msun$\label{tbl:comp}}
\tabletypesize{\scriptsize}
\tablewidth{0pt}
\tablehead{
\multicolumn{3}{c}{Parameter Combinations} &
\multicolumn{3}{c}{Mass Fractions} &
\multicolumn{3}{c}{Isotopic Ratios} \\
\colhead{$M_{WD}$} & \colhead{$T_{WD}$} & \colhead{$\dot{M}$} &
\colhead{$^{12}$C} & \colhead{$^{14}$N} & \colhead{$^{16}$O} &
\colhead{$^{13}$C/$^{12}$C} & \colhead{$^{15}$N/$^{14}$N} &
\colhead{$^{17}$O/$^{16}$O} \\
}
\startdata
0.50 & 4 & 5E-11 & 7.8E-3 & 4.8E-2 & 5.7E-2 & 6.1E-1 & 2.6E-4 & 3.5E-2 \\
0.50 & 3 & 5E-11 & 7.1E-3 & 4.0E-2 & 4.8E-2 & 5.9E-1 & 2.0E-4 & 5.2E-2 \\
0.50 & 2 & 7E-11 & 1.1E-2 & 5.2E-2 & 6.3E-2 & 4.1E-1 & 6.8E-5 & 3.0E-2 \\
\hline
0.50 & 3 & 1E-10 & 5.5E-3 & 6.1E-2 & 6.6E-2 & 7.4E-1 & 3.5E-5 & 2.0E-2 \\
\enddata
\end{deluxetable}

\begin{deluxetable}{ccccccccccc}
\tablecaption{Characteristics of The Outburst -- Very low $T_{WD}$\label{tbl:cold}}
\tabletypesize{\scriptsize}
\tablewidth{0pt}
\tablehead{
\multicolumn{3}{c}{Parameter Combinations} &
\multicolumn{7}{c}{Outburst Characteristics} \\
\colhead{$M_{WD}$} & \colhead{$T_{WD}$} & \colhead{log$\dot{M}$} &
\colhead{$m_{acc}$} & \colhead{$m_{ej}$} & \colhead{$Z_{ej}$} &
\colhead{$v_{av}$} &
\colhead{$T_{8,max}$} & \colhead{$L_{4,max}$} & \colhead{$t_{3,bol}$} \\
\colhead{($\Msun$)} & \colhead{($10^{6}$~K)} & \colhead{($\Msun$~yr$^{-1}$)} &
\colhead{($\Msun$)} & \colhead{($\Msun$)} & \colhead{} &
\colhead{$(km~s^{-1})$} &
\colhead{($10^{8}$~K)} & \colhead{$(10^{4}~L_\odot)$} & \colhead{(days)}\\
}
\startdata
0.40 & 1.79 & -10 & 9.2E-4 & 1.0E-3 & 0.15 & 147 & 1.30 & ~9.4 & 1.7E3 \\
0.50 & 1.71 & -10 & 6.5E-4 & 7.9E-4 & 0.21 & 480 & 1.30 & 17.9 & 1.1E4 \\
0.65 & 1.65 & -10 & 4.7E-4 & 5.3E-4 & 0.15 & 424 & 1.83 & 29.3 & 1.9E3 \\
\enddata
\end{deluxetable}

\begin{figure}
\figurenum{1a}
\epsscale{0.6}
\plotone{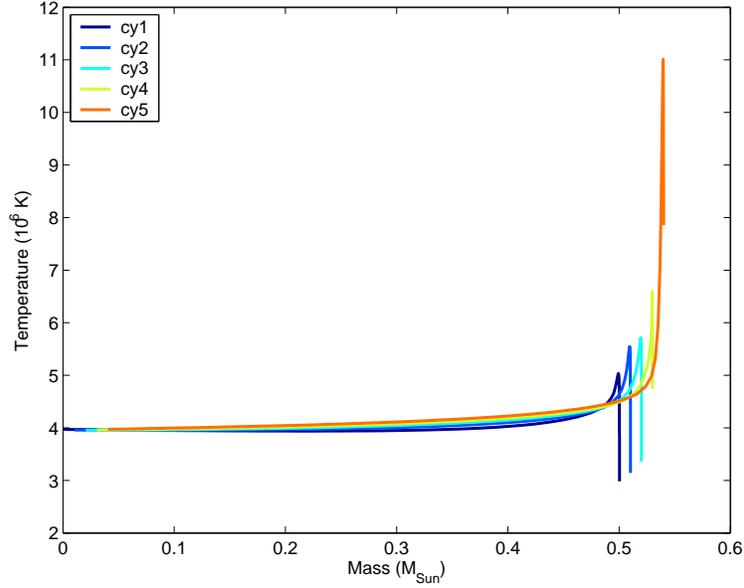}
\caption{Temperature profiles at the beginning of the accretion phase for 5 consecutive cycles for a given WD and accretion rate, as obtained in the present calculations (first model in Table \ref{tbl:charact}). The profiles in each panel are slightly shifted in mass with respect to one another, in order to enable the comparison.}
\label{fig:twd1}
\end{figure}

\begin{figure}
\figurenum{1b}
\plotone{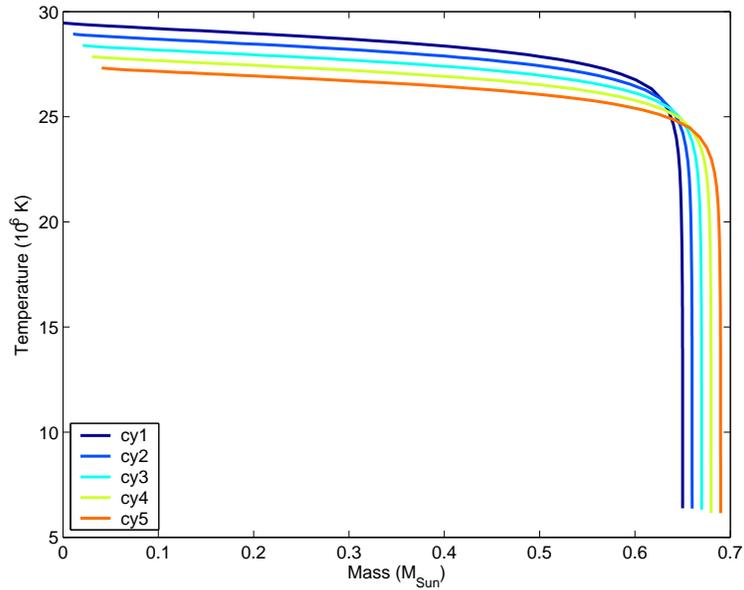}
\caption{Temperature profiles at the beginning of the accretion phase for 5 consecutive cycles for a given WD and accretion rate, as obtained in Paper II, for a $0.65\Msun$ WD of temperature $30\times10^{6}$~K, which has accreted mass at a rate of $1\times10^{-11}\Msun$~yr$^{-1}$. The profiles in each panel are slightly shifted in mass with respect to one another, in order to enable the comparison.}
\label{fig:twd2}
\end{figure}

\begin{figure}
\figurenum{2}
\epsscale{0.5}
\plotone{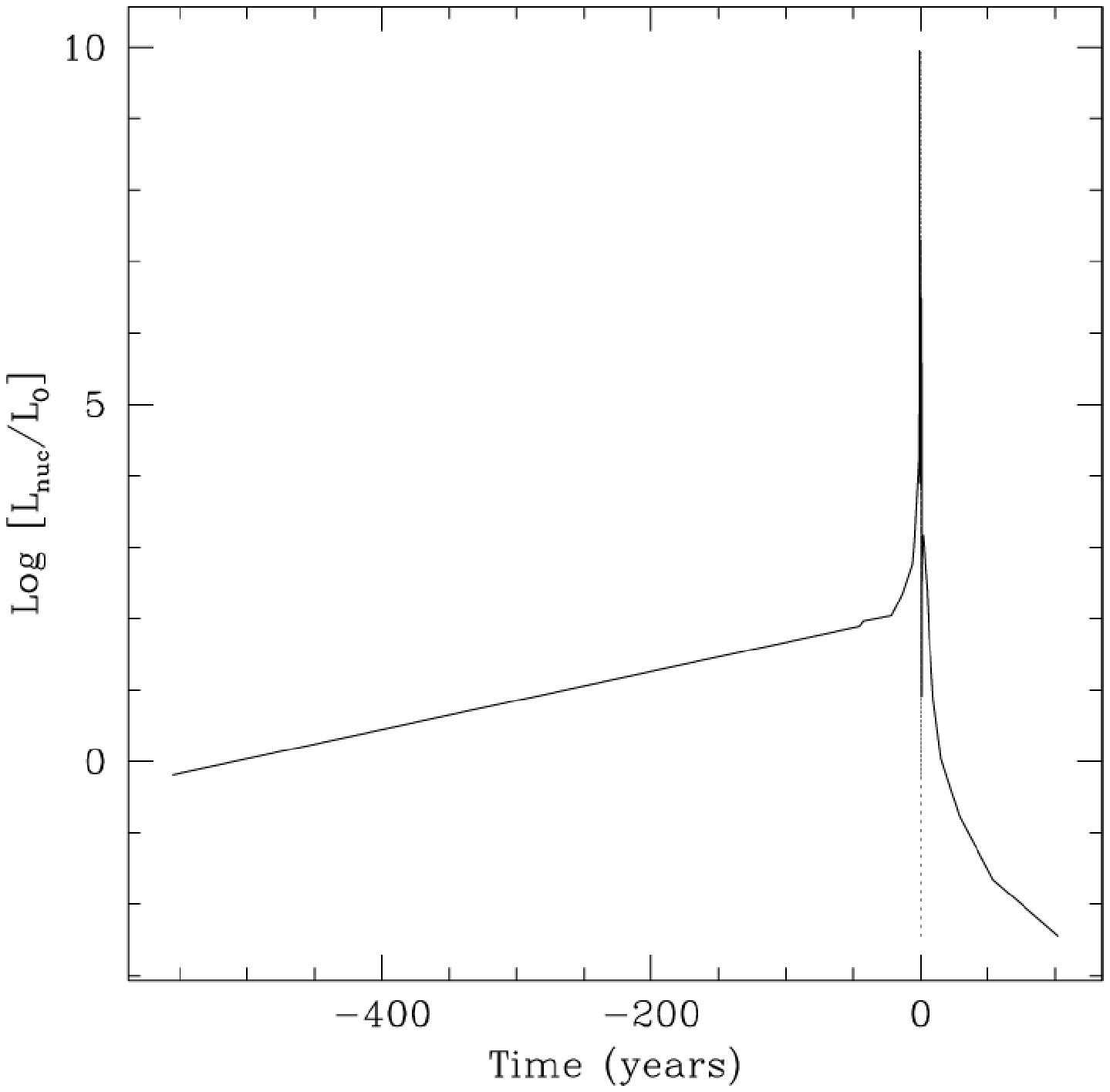}
\caption{Nuclear luminosity history for a $0.5\Msun$ WD of temperature $2\times10^{6}$~K, which has accreted mass at a rate of $7\times10^{-11}\Msun$~yr$^{-1}$, in the centuries before and after eruption}
\label{fig:twd2}
\end{figure}

\begin{figure}
\figurenum{3}
\plotone{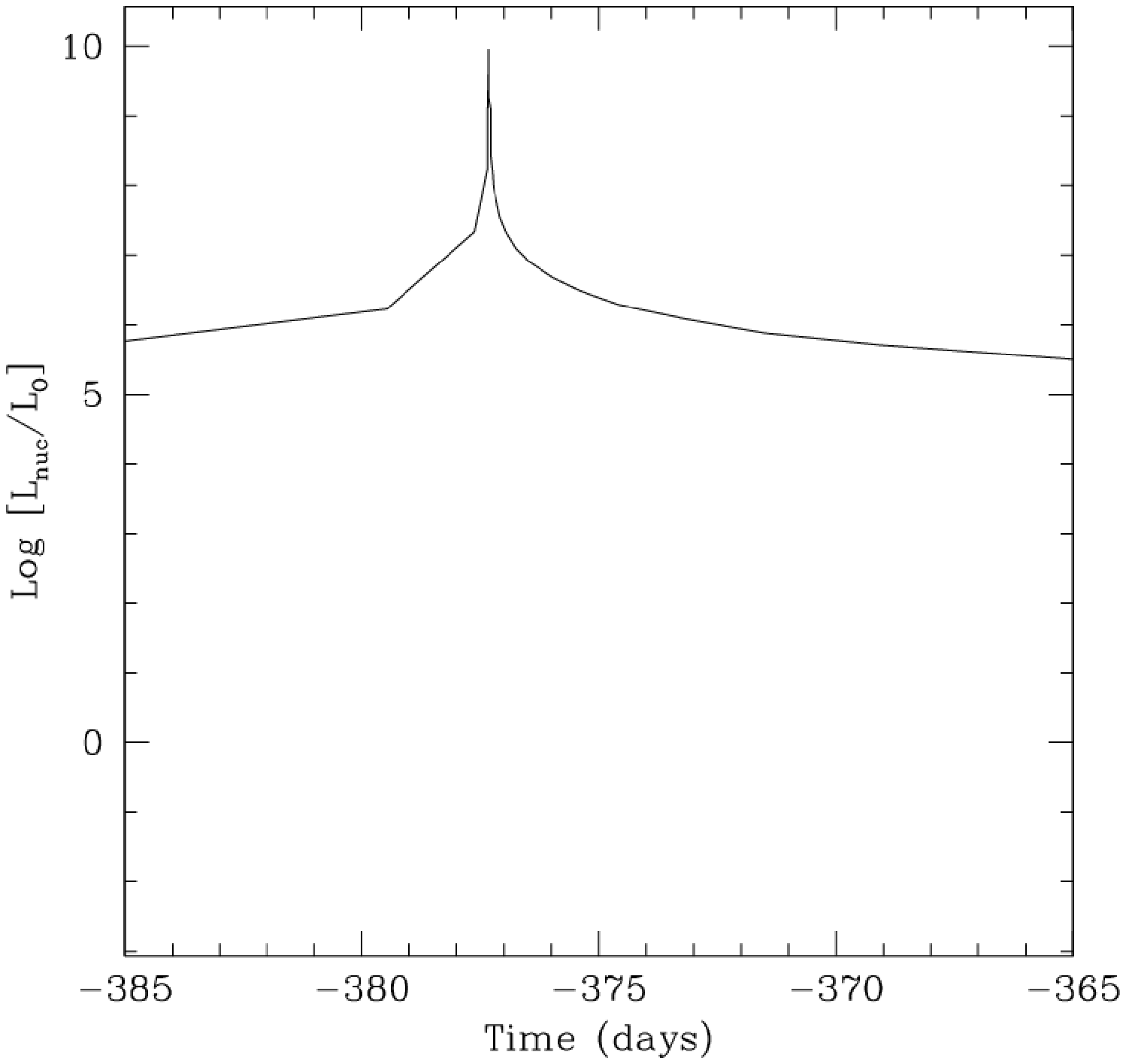}
\caption{The luminosity generated in the first thermonuclear pulse of a $0.5\Msun$ WD of temperature $2\times10^{6}$~K, which has accreted mass at a rate of $7\times10^{-11}\Msun$~yr$^{-1}$, a year before the optical outburst begins}
\label{fig:earlypulselum}
\end{figure}

\begin{figure}
\figurenum{4}
\epsscale{1.1}
\plottwo{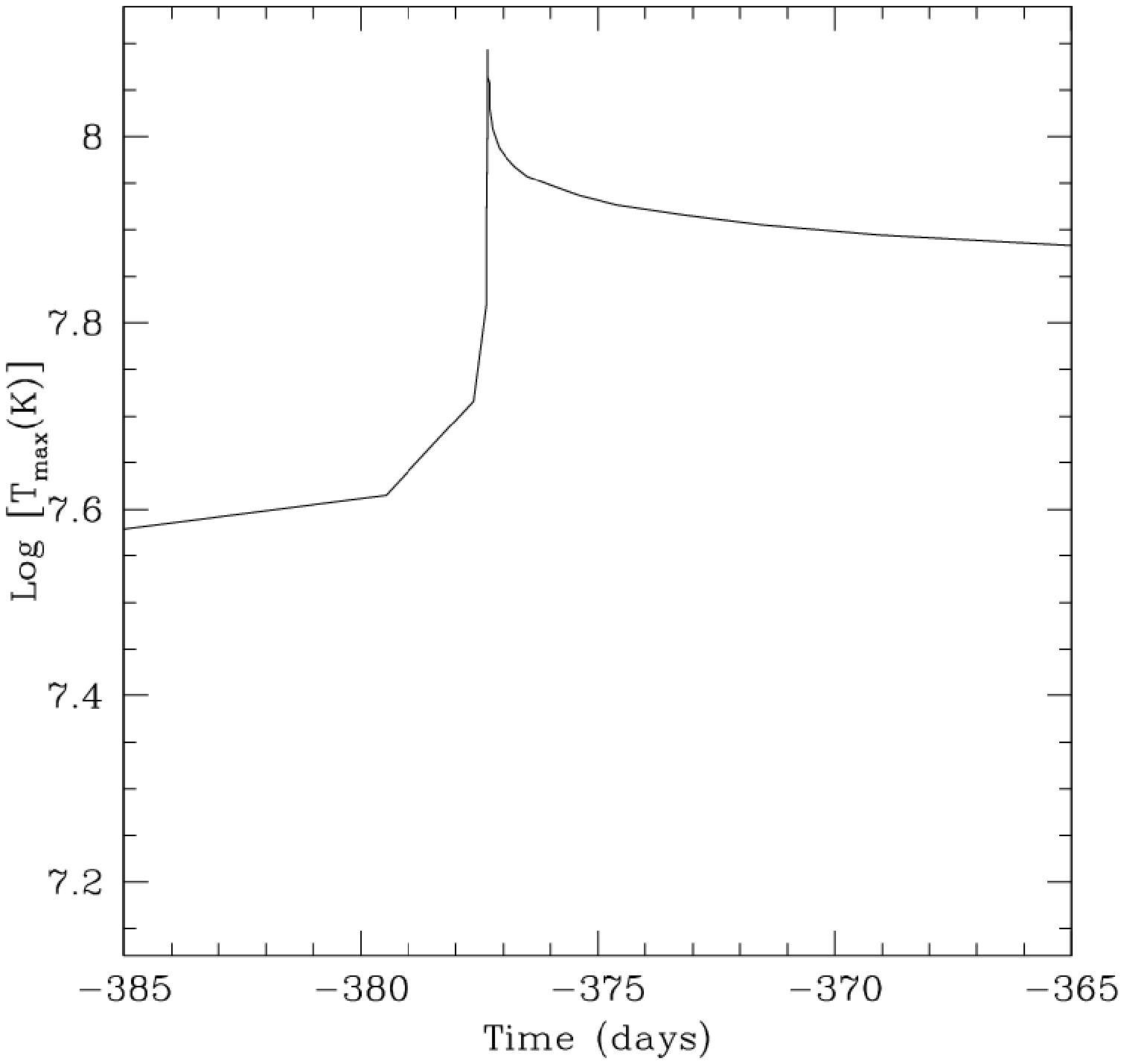}{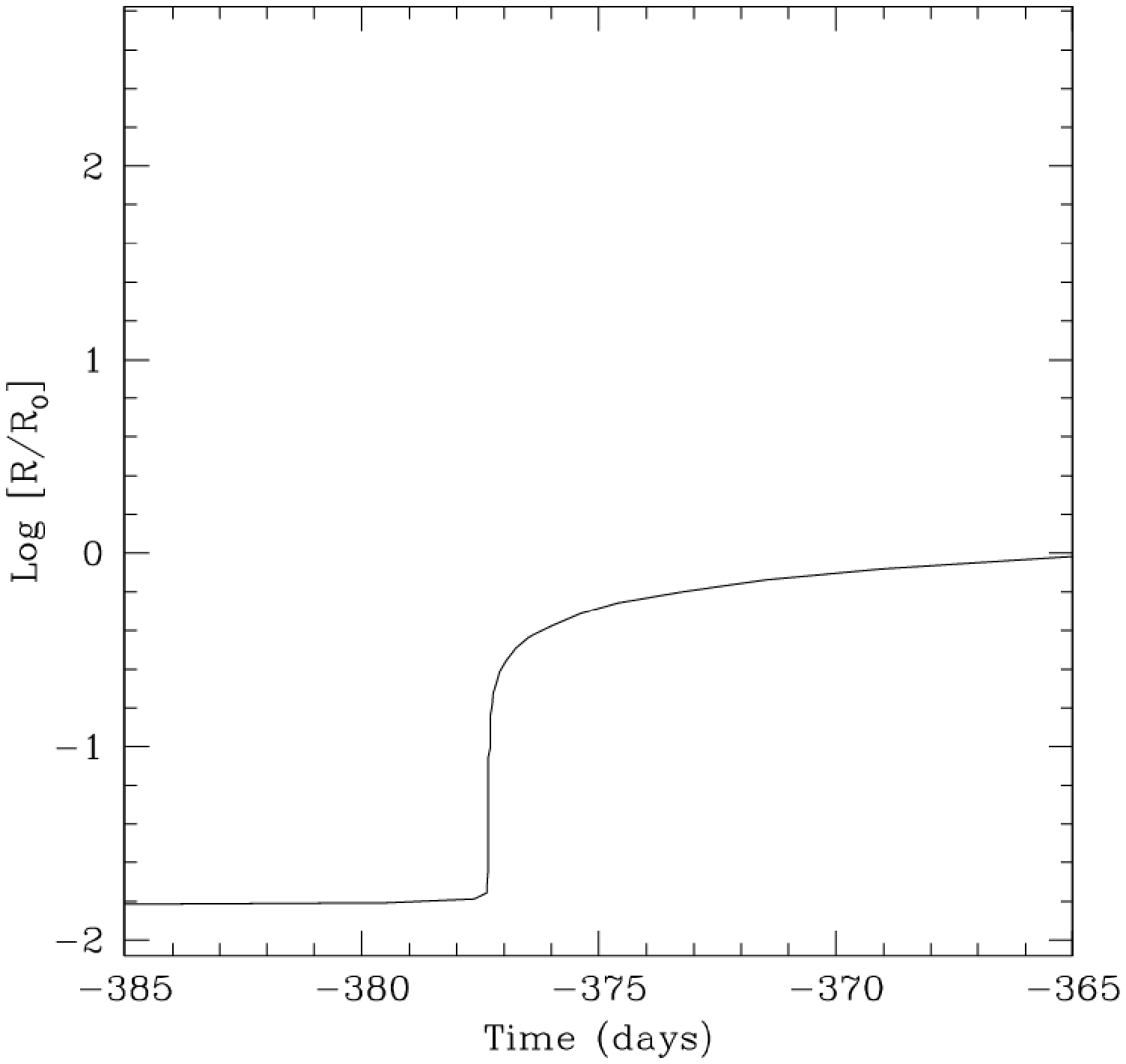}
\plottwo{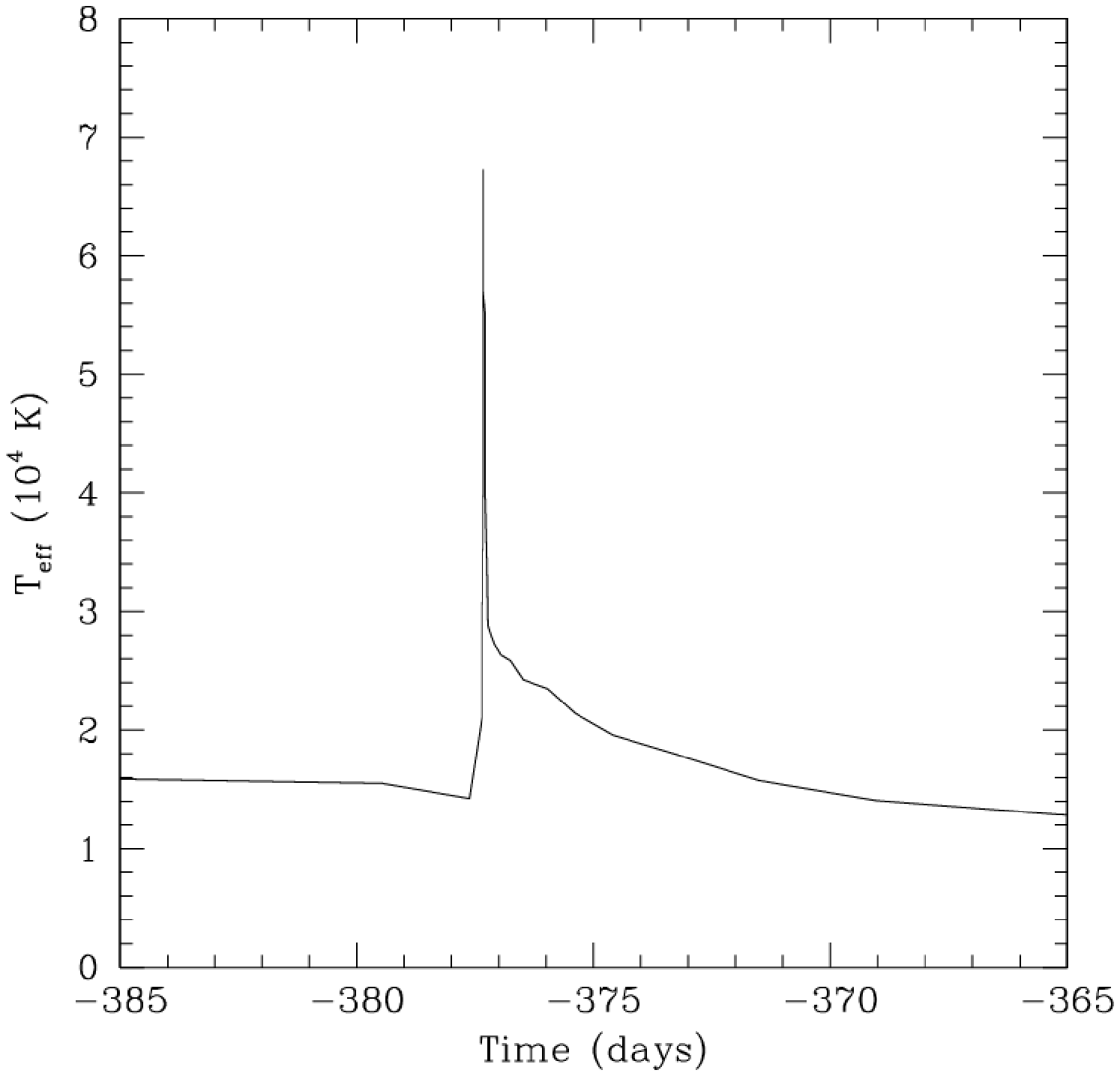}{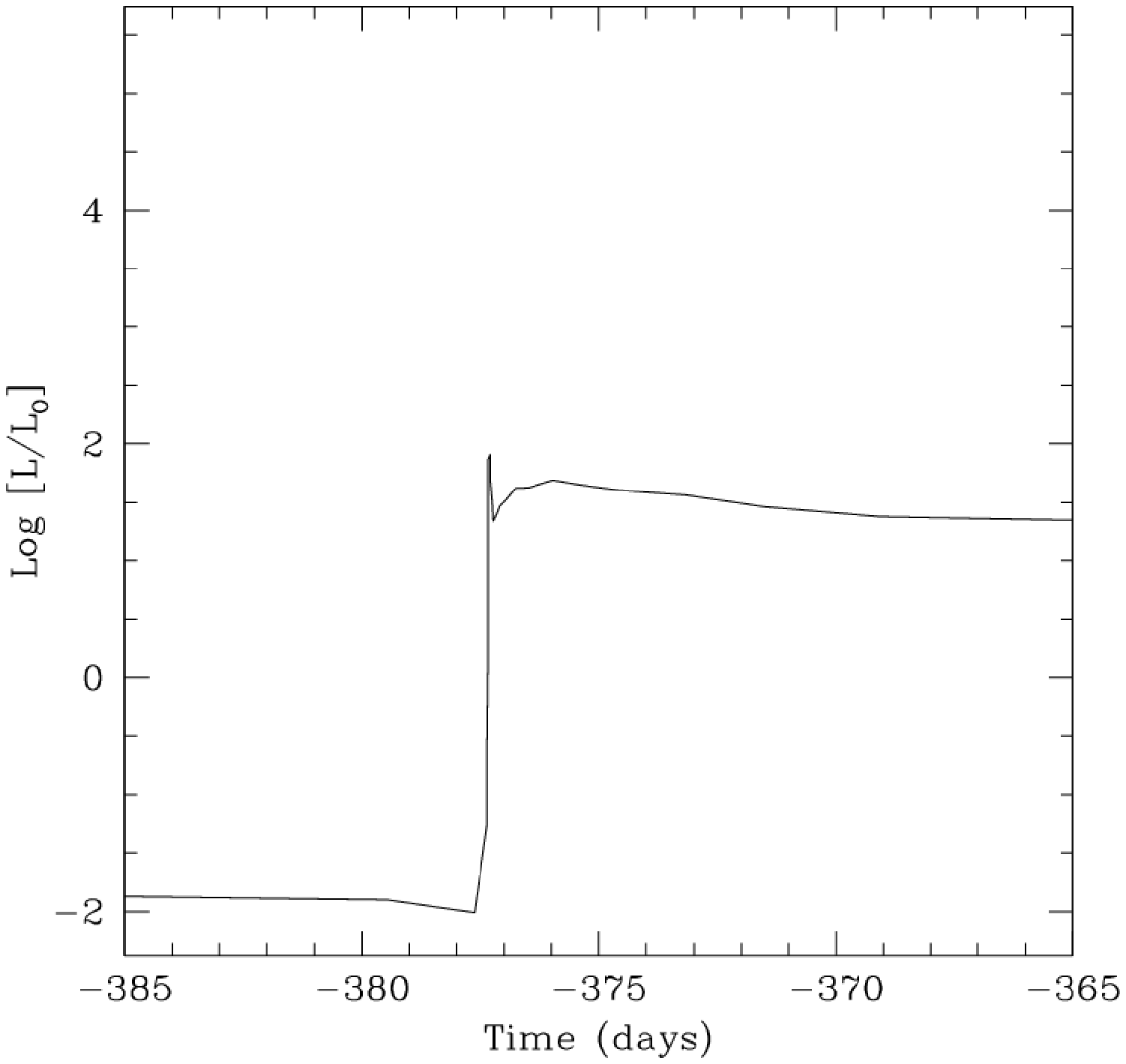}
\caption{The first outburst: pre-mass loss phase of an outburst cycle obtained for a $0.5\Msun$ WD of temperature $2\times10^{6}$~K, which has accreted mass at a rate of $7\times10^{-11}\Msun$~yr$^{-1}$: {\it top left} The temperature history at the envelope base of the first thermonuclear pulse, a year before the major outburst begins; {\it top right} radius of the WD envelope; {\it bottom left} envelope effective temperature; {\it bottom right} envelope apparent luminosity.}
\label{fig:earlypulsehistory}
\end{figure}

\begin{figure}
\figurenum{5}
\epsscale{1.1}
\plottwo{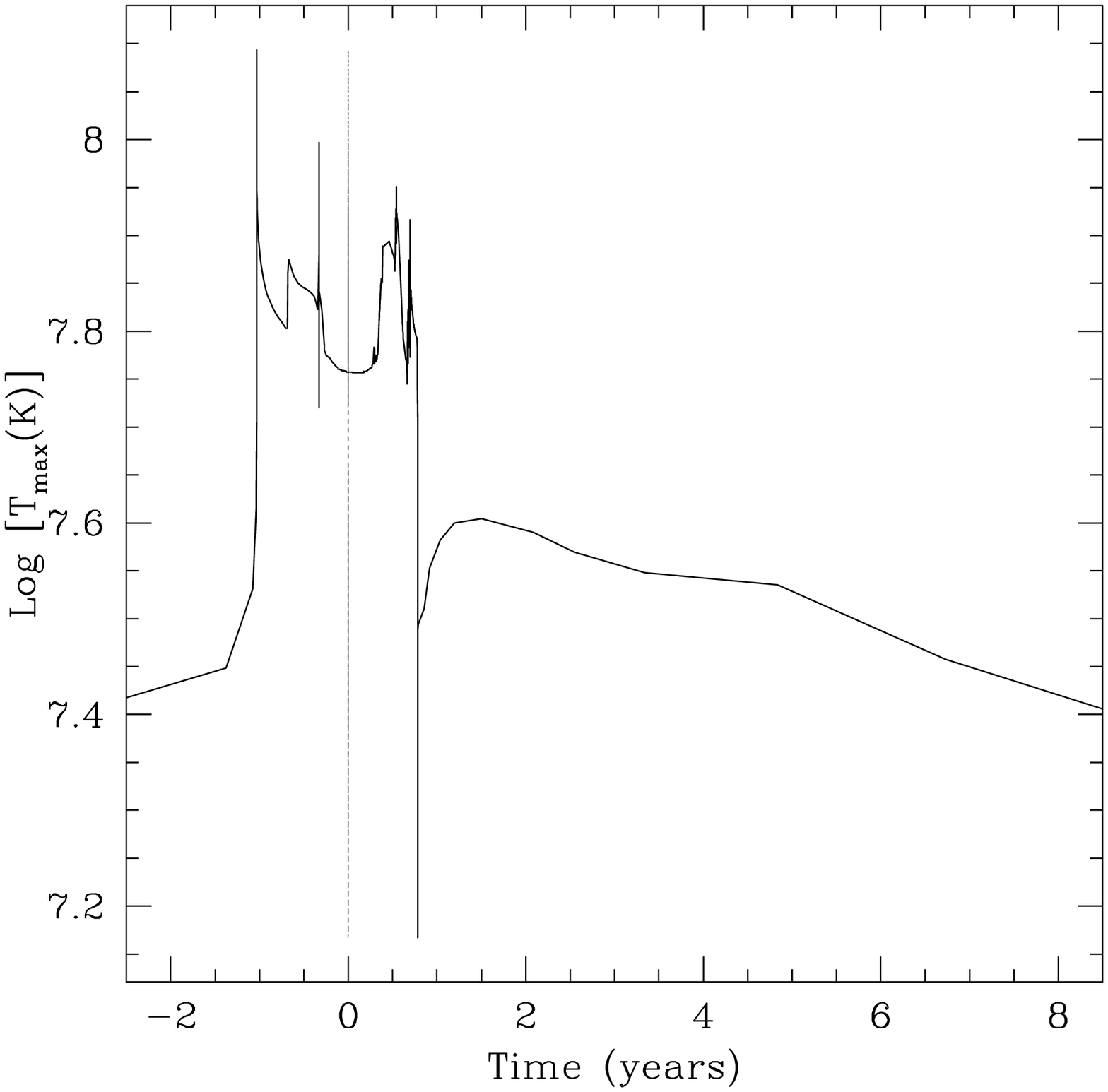}{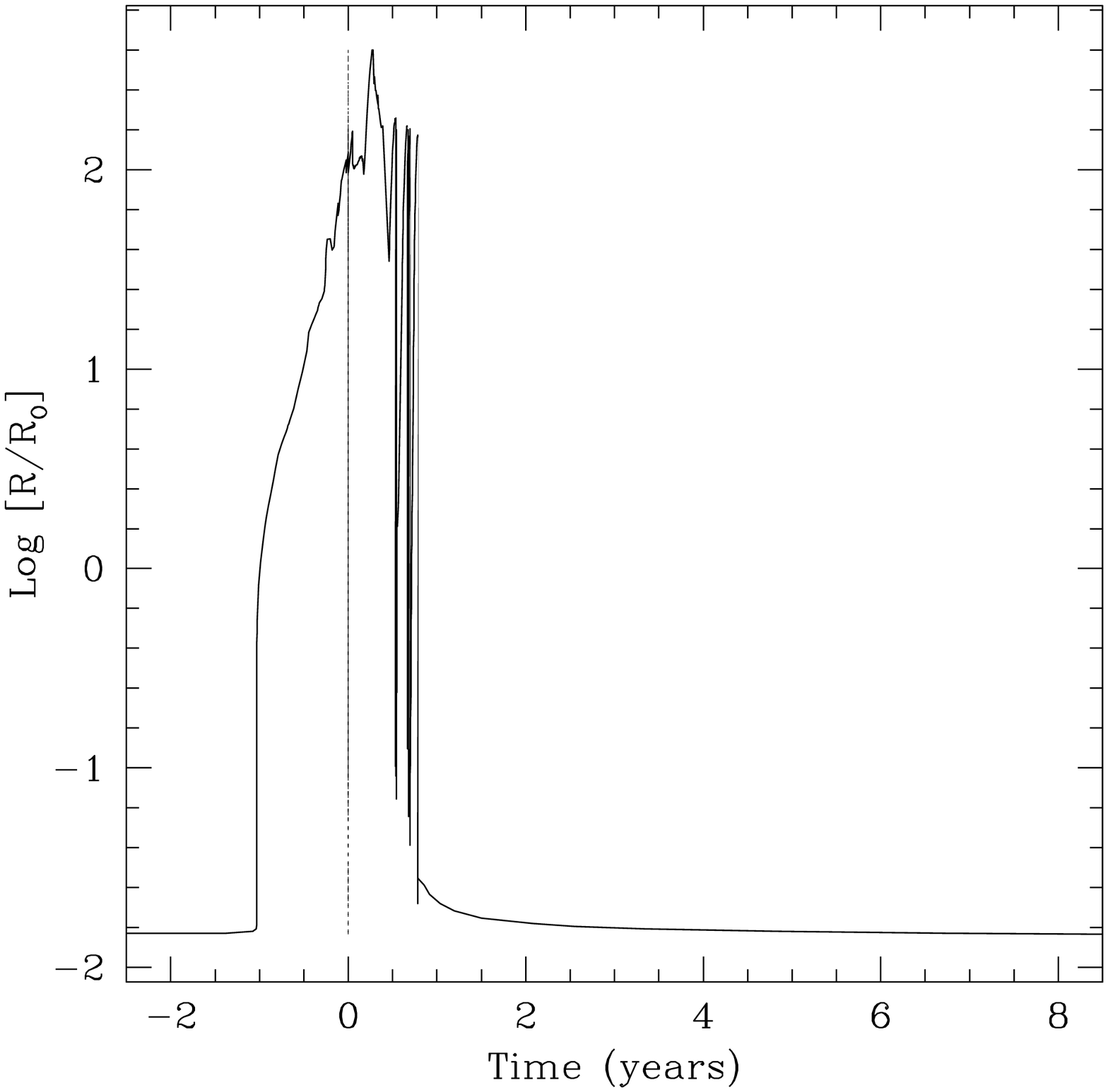}
\plottwo{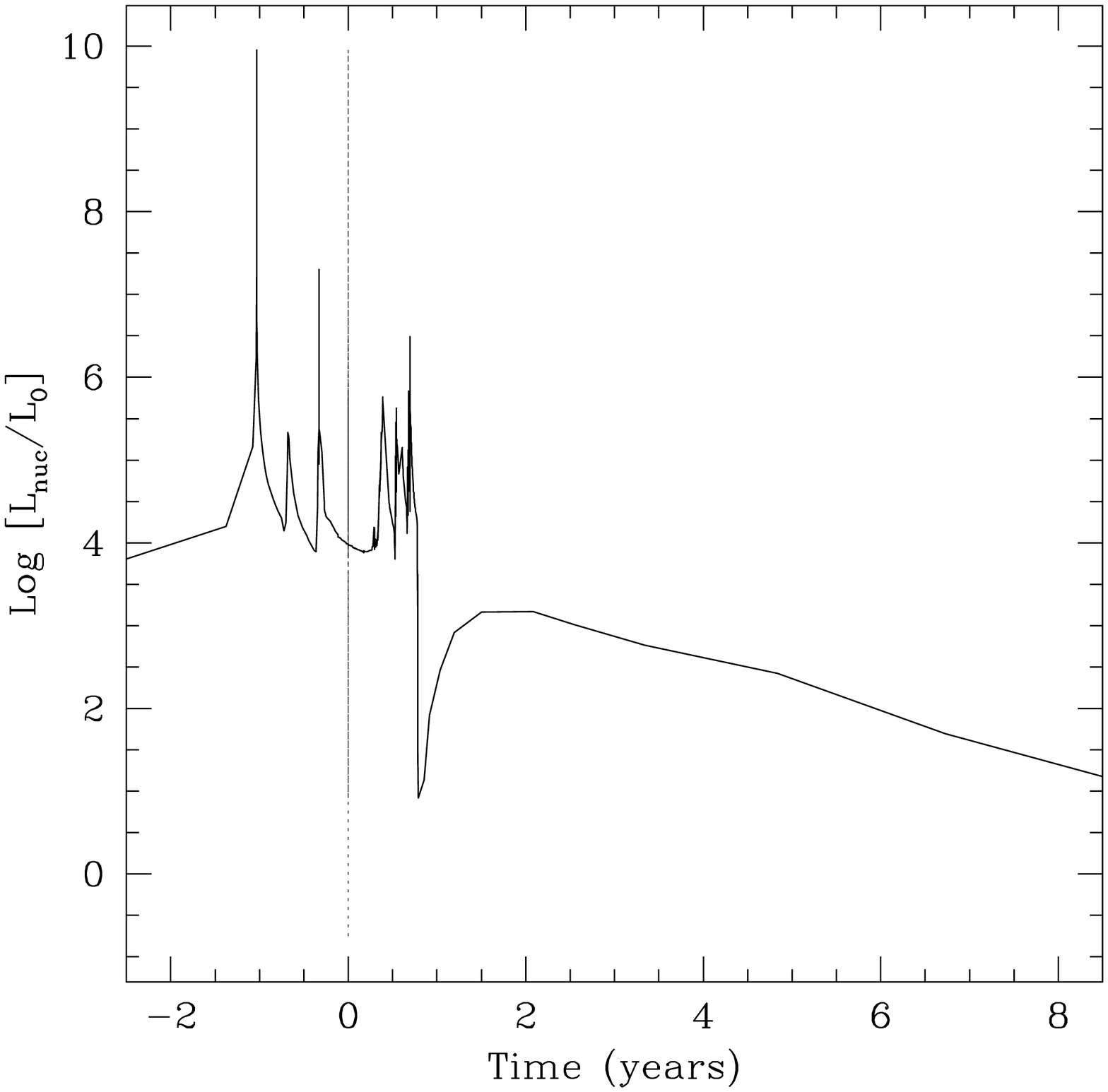}{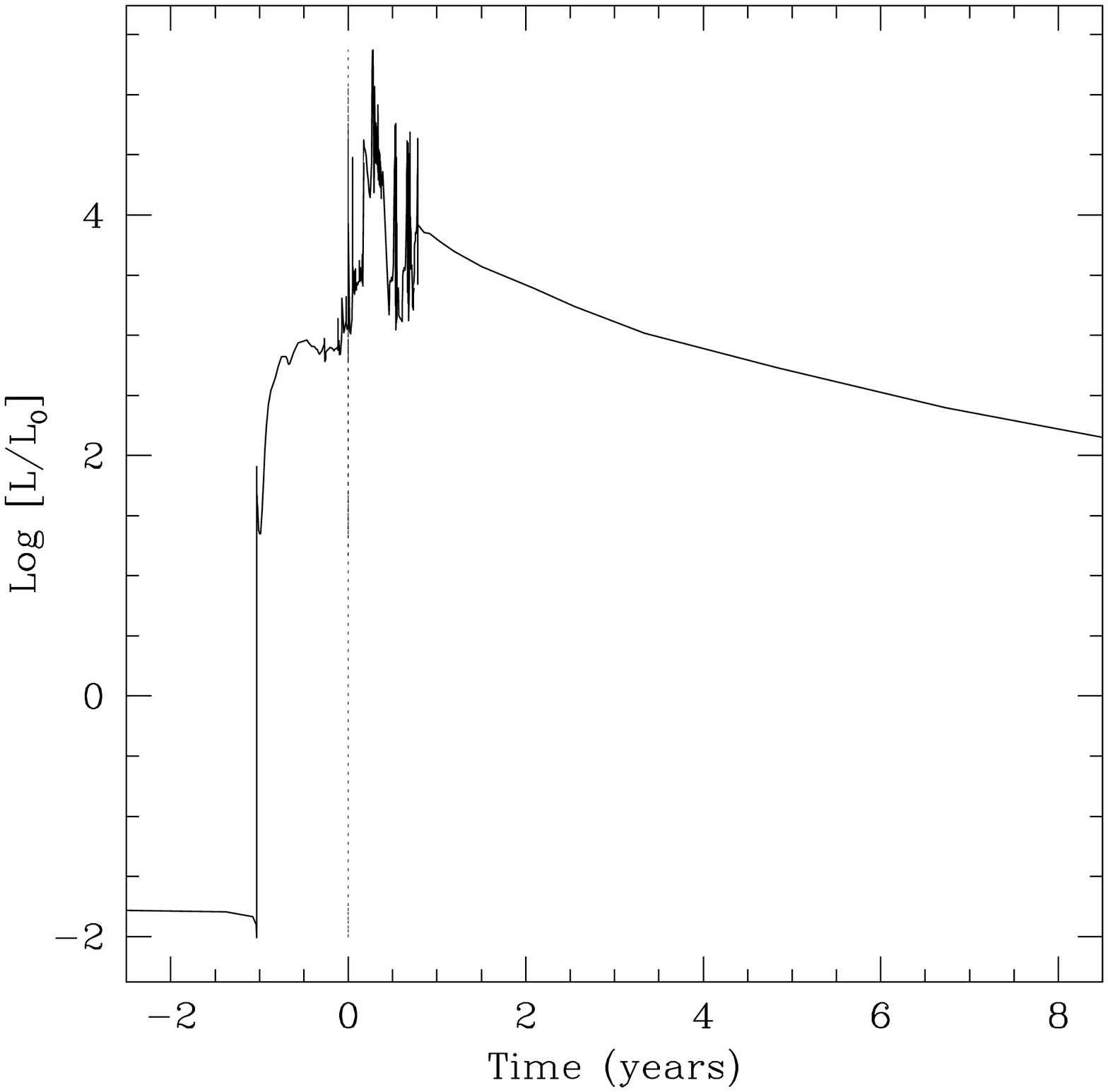}
\caption{The outburst cycle obtained for a $0.5\Msun$ WD of temperature $2\times10^{6}$~K, which has accreted mass at a rate of $7\times10^{-11}\Msun$~yr$^{-1}$: {\it top left}
The temperature history at the envelope base; {\it top right} radius of the WD envelope; {\it bottom left} six nuclear luminosity pulses generated in the white dwarf envelope ; {\it bottom right} envelope apparent luminosity. This peak apparent luminosity is artificially limited by the opacities used in the simulations, and can be 10 to 100 times larger when the envelope effective temperature is 3000 - 4000 Kelvins.}
\label{fig:outbursthistory}
\end{figure}

\begin{figure}
\figurenum{6}
\epsscale{1.1}
\plotone{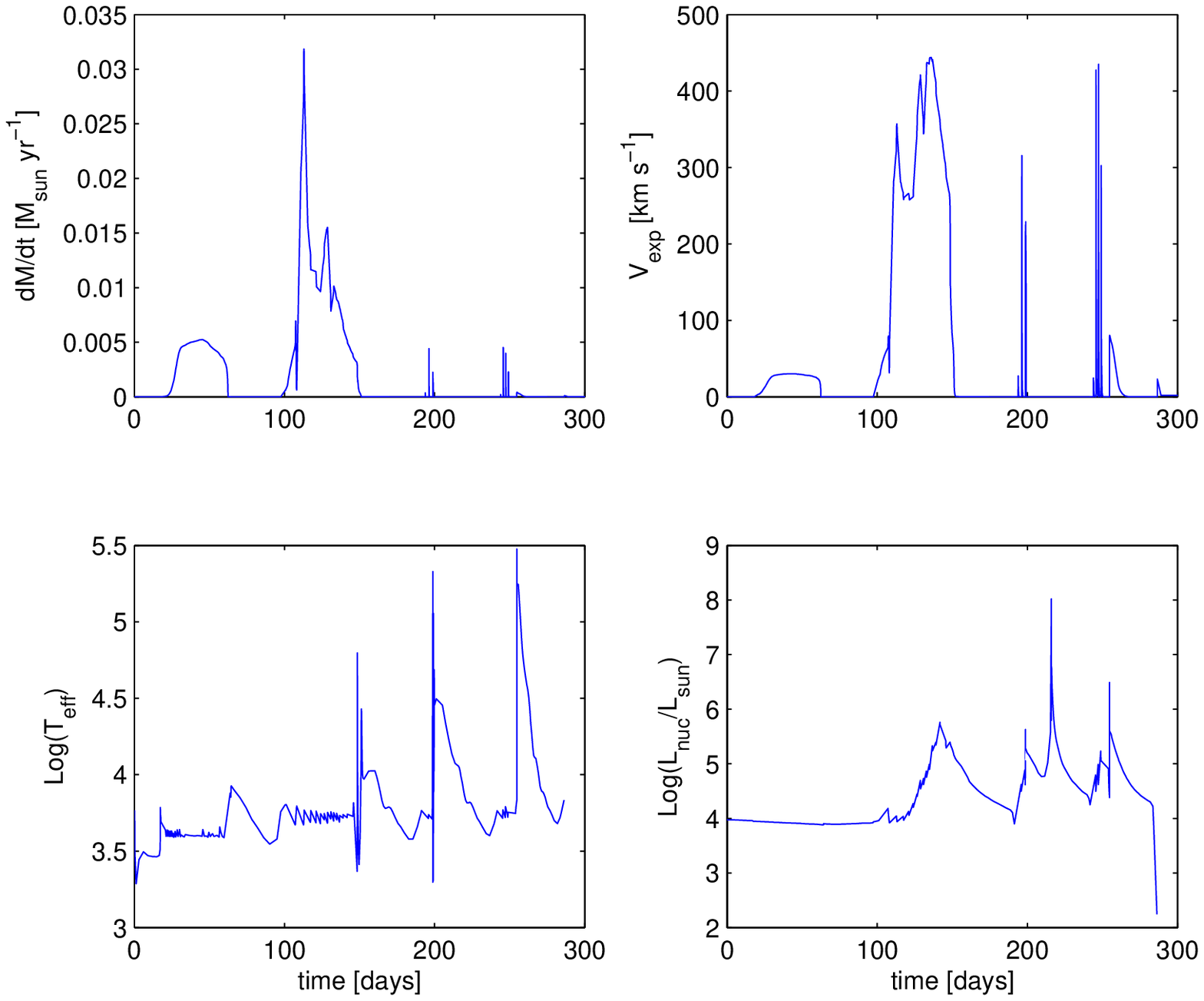}
\caption{The mass loss phase of an outburst cycle obtained for a $0.5\Msun$ WD of temperature $2\times10^{6}$~K, which has accreted mass at a rate of $7\times10^{-11}\Msun$~yr$^{-1}$: {\it top left}
mass loss rate; {\it top right} expansion velocity; {\it bottom left} effective temperature; {\it bottom right} nuclear luminosity.}
\label{fig:evol}
\end{figure}

\begin{figure}
\figurenum{7}
\epsscale{1.0}
\plotone{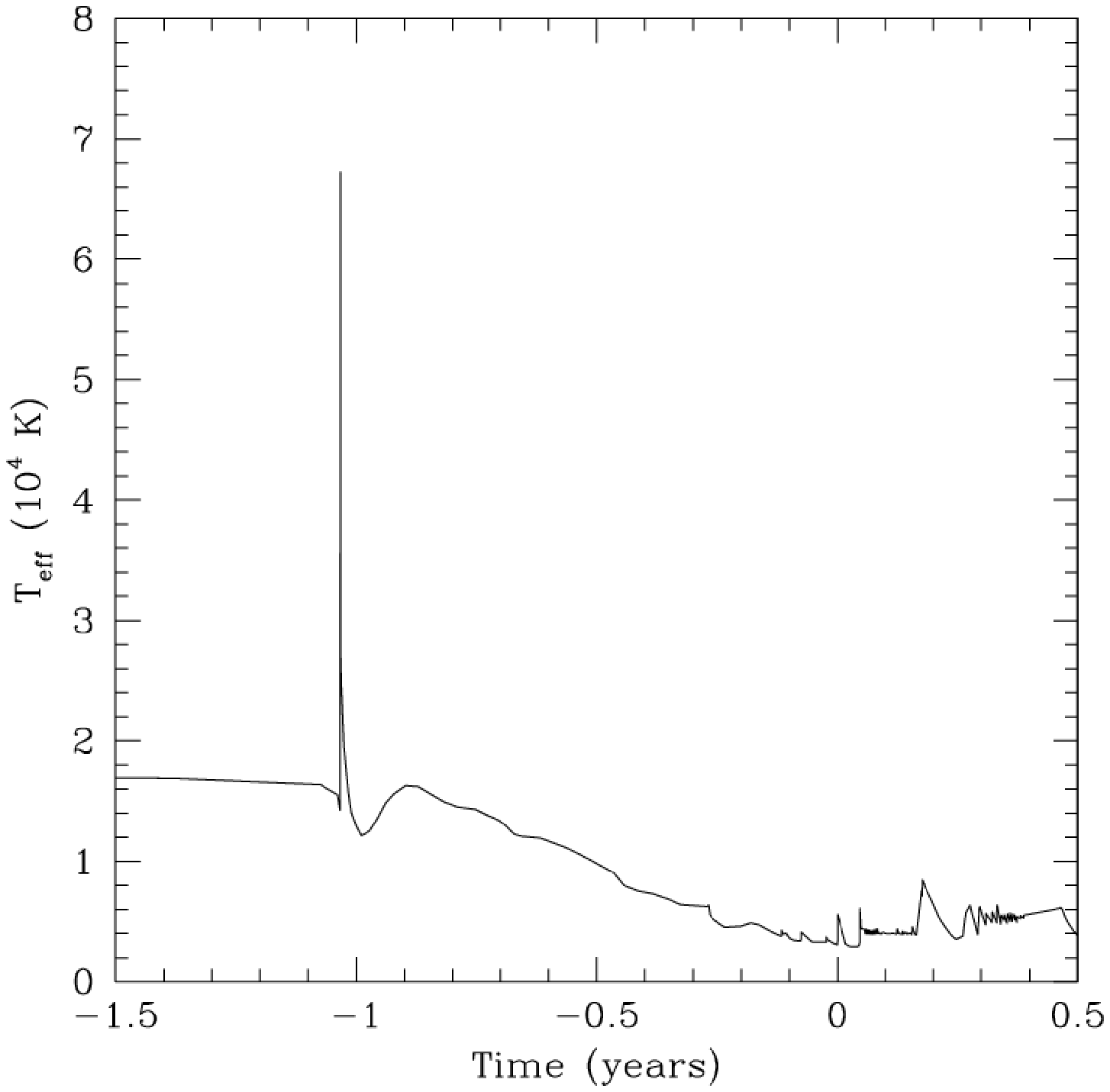}
\caption{The outburst effective temperature obtained for a $0.5\Msun$ WD of temperature $2\times10^{6}$~K, which has accreted mass at a rate of $7\times10^{-11}\Msun$~yr$^{-1}$.
The model is the same as that of Fig.~\ref{fig:evol} and time is set to zero at the same point.}
\label{fig:Teff}
\end{figure}

\end{document}